\documentclass[journal=jacsat,manuscript=article,layout=twocolumn]{achemso}
\setkeys{acs}{doi = true}

\usepackage{graphicx}
\usepackage{dcolumn}
\usepackage{bm}
\usepackage[hidelinks]{hyperref}
\usepackage{orcidlink}
\usepackage[normalem]{ulem}
\usepackage{float}
\usepackage{booktabs}
\usepackage{amsmath}
\usepackage{amssymb}
\usepackage{soul}
\usepackage{makecell}

\DeclareRobustCommand{\doi}[1]{%
    \begingroup
    \urlstyle{tt}%
    \href{https://doi.org/#1}{\uline{\nolinkurl{#1}}}%
    \endgroup
}

\usepackage{color}
\usepackage{xcolor}

\usepackage{enumitem}
\setlist[itemize]{leftmargin=*}

\usepackage[
    group-separator={\,},
    separate-uncertainty=true,
    bracket-ambiguous-numbers=false,
    retain-zero-uncertainty=true,
    multi-part-units=single,
    list-units=single,
    range-units=single,
]{siunitx}  
\DeclareSIUnit[per-mode=symbol]{\gpcc}{\gram\per\cubic\centi\metre}  
\DeclareSIUnit{\angstrom}{\text{Å}}  
\DeclareSIUnit[per-mode=symbol]{\ev}{\electronvolt}  
\DeclareSIUnit[per-mode=symbol]{\evpang}{\electronvolt\per\angstrom}  
\DeclareSIUnit[per-mode=symbol]{\mevpang}{\milli\electronvolt\per\angstrom}  
\DeclareSIUnit{\atom}{\text{atom}}  
\DeclareSIUnit[per-mode=symbol]{\mevpatom}{\milli\electronvolt\per\atom}  
\DeclareSIUnit{\cal}{\text{cal}}  
\DeclareSIUnit[per-mode=symbol]{\kcalpermol}{\kilo\cal\per\mol}  

\newcolumntype{d}{D{.}{.}{4.2}}

\title{Understanding multi-fidelity training of machine-learned force-fields}

\author{John L.A. Gardner\,\orcidlink{0009-0006-7377-7146}}
\altaffiliation{Work undertaken while at Microsoft Research, AI for Science.}
\affiliation{Department of Chemistry, Inorganic Chemistry Laboratory,\\University of Oxford, Oxford OX1\,3QR, UK}
\author{Hannes Schulz\,\orcidlink{0000-0001-6408-9794}}
\author{Jean H\'{e}lie\,\orcidlink{0000-0003-3413-6865}}
\author{Lixin Sun\,\orcidlink{0000-0002-7971-5222}}
\author{Gregor N.C. Simm\,\orcidlink{0000-0001-6815-352X}}
\email{gregorsimm@microsoft.com}
\affiliation{Microsoft Research, AI for Science,\\21 Station Road, Cambridge CB1\,2FB, UK}

\begin{document}

\begin{abstract}
    This study systematically investigates two multi-fidelity strategies used to train machine-learned force fields (MLFFs)---pre-training/fine-tuning and multi-headed training---and elucidates the mechanisms underpinning their success.
    For pre-training and fine-tuning, we uncover a log-log linear relationship between pre-trained and fine-tuned accuracies that holds across model architectures, model sizes, and quantum-chemical methods.
    The success of this approach hinges on the quantity and quality of available pre-training data, and, critically, the inclusion of force labels.
    We demonstrate that pre-trained representations are inherently method-specific, requiring adaptation of the model backbone during fine-tuning.
    In contrast, multi-headed models learn method-independent backbone representations, where again the heads' accuracies are log-log linearly related.
    Relative to pre-training and fine-tuning, these shared representations marginally reduce model performance in most cases.
    However, this trade-off is offset by practical advantages: multi-headed training extends naturally to multiple labelling methods and enables partial replacement of expensive labels with cheaper alternatives, paving the way towards cost-efficient universal MLFFs.
\end{abstract}

\maketitle

\section{Introduction}

Machine-learned force fields (MLFFs)\cite{Unke2021Machine,Behler2021Four,Friederich2021Machinelearned,Deringer2019MLIPs} are transforming atomistic simulations by predicting quantum-chemical properties at a fraction of the computational cost of conventional \textit{ab initio} methods, such as density functional theory (DFT)\cite{Hohenberg-64-11,Kohn-65-11} or coupled cluster (CC)\cite{Purvis-82-02}.
Their applications span diverse fields, including materials discovery and drug design.\cite{Deringer-20-10a,Zhou-23-10,Yang-24-05,Chen-24-03}
The community is increasingly working towards \textit{universal} force fields: single models capable of accurate predictions across broad regions of chemical space.\cite{Shoghi-23-10,Batatia2025Foundation,Yang-24-05,Rhodes2025Orbv3,Wood2026UMA}
Building such models requires large volumes of diverse and accurate training data, the generation of which poses two distinct challenges.

First, the most accurate quantum-chemical methods, notably coupled-cluster theory at the CCSD(T) level\cite{Purvis-82-02}, scale steeply with system size, making it prohibitively expensive to generate all training data at high fidelity.
Second, no single quantum-chemical method is universally the most applicable across all of chemical space: CCSD(T) is widely regarded as the gold standard for molecular systems,\cite{Smith-20-05} while periodic DFT and multi-reference methods may be more suitable for the description of inorganic crystals or strongly correlated systems, respectively.
Together, these challenges motivate \textit{multi-fidelity} training strategies that leverage data from multiple quantum-chemical methods (or labelling methods), each of which strikes a different balance between accuracy and computational cost.
There is growing evidence, both in broader machine learning\cite{Zhuang-21-01} and in the MLFF community,\cite{Smith-19-07,Gardner-24-03,Cui-24-11} that training on abundant, low-fidelity data yields representations that transfer usefully to a high-fidelity target task---a phenomenon termed \textit{positive transfer}.

\begin{figure}[ht]
    \centering
    \includegraphics[width=\linewidth]{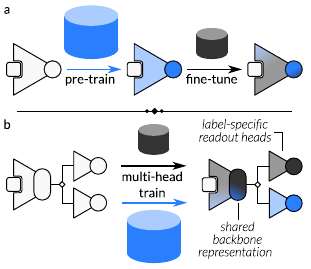}
    \caption{\label{fig:overview}%
        Comparison of two multi-fidelity training strategies for MLFFs considered in this work.
        \textbf{(a)} A sequential pre-training and fine-tuning approach. The model is first trained on a large dataset of low-fidelity labels (blue, e.g., DFT) and subsequently fine-tuned on a smaller dataset of high-fidelity labels (black, e.g., CC).
        \textbf{(b)} A multi-headed approach where a shared model backbone learns from multiple label fidelities simultaneously, each through a dedicated readout head.
    }
\end{figure}

Multi-fidelity strategies for MLFFs broadly fall into two categories.
The first, pre-training and fine-tuning, initially trains a model on abundant, computationally inexpensive labels before fine-tuning on a smaller set of high-fidelity labels (Fig.~\ref{fig:overview}a).
This sequential approach has demonstrated consistent positive transfer across a range of labelling methods and architectures.\cite{Smith-19-07,Shui-22-10,Gardner-23-06,Gardner-24-03,Cui-24-11}
The second category involves architectural modifications that enable a single model to learn from multiple labelling methods concurrently (Fig.~\ref{fig:overview}b).
This approach facilitates the integration of several diverse datasets, each labelled by a different method, and is growing in popularity as more datasets become publicly available.\cite{Egorova-20-10,Shoghi-23-10,Thaler-24-12,Zhang-24-12,Chen-25-04,Huang-25-04,Levine-25-05,Batatia-2025-CrossLearning}

While multi-fidelity training is well understood in broader ML contexts,\cite{Kennedy-00,Forrester-07-10,Huang-22-04,Mikkola-23-04} the mechanisms driving positive transfer in the MLFF domain remain largely unexplored.
Rather than proposing new multi-fidelity strategies, this work investigates the mechanisms underpinning the success of existing strategies.
Through systematic ablation studies on the pre-train/fine-tune strategy, we identify the mechanisms responsible for positive transfer from low- to high-fidelity labels and establish a quantitative relationship between pre-training and fine-tuning accuracy.
We then turn to multi-headed training and uncover striking mechanistic parallels with the sequential approach.
We conclude with practical recommendations for multi-fidelity training of MLFFs.

\section{Related Work}
\label{sec:related-work}

Initial MLFF approaches involved hand-crafted descriptors processed by multi-layer perceptrons,\cite{Behler-07-04} Gaussian processes,\cite{Bartok-10-04} or linear models.\cite{Drautz-19-01}
Subsequently, models based on graph neural network (GNN) architectures have surpassed descriptor-based models in accuracy on most benchmarks.\cite{Schutt-18-03,Gasteiger-22-04a,Gasteiger-22-04b,Batzner-22-05,Batatia-22,Musaelian-22-04,Wang2024Enhancing,Liao2024EquiformerV2,Rhodes2025Orbv3,Wood2026UMA}
In this study, we compare two state-of-the-art GNNs: MACE \cite{Batatia-22} and Allegro \cite{Musaelian-22-04}.
Both architectures employ spherical tensor decompositions to generate equivariant representations of the local chemical environment.
MACE explicitly constructs many-body features in each layer, whereas Allegro enforces locality by using local edge convolutions to incorporate many-body information progressively over multiple layers.

In machine learning, multi-fidelity training refers to strategies where a model learns from multiple datasets that contain labels of the same underlying property but at varying levels of accuracy, where higher accuracy typically incurs greater computational cost.
This contrasts with multi-task training (a different flavour of transfer learning) where a model learns to predict several, potentially unrelated properties.
This study focuses exclusively on multi-fidelity training strategies for MLFFs.
We briefly review relevant prior work below.

\subsection{Pre-training and Fine-tuning}

Pre-training equips models with transferable representations learned from auxiliary data, enabling higher accuracy when fine-tuning on a specific, data-limited task than training from scratch.\cite{Chen-20-02,Mikolov-13-01,Peters-18-02}
In the context of MLFFs, pre-training on abundant, low-fidelity data followed by fine-tuning on scarce, high-fidelity data has consistently demonstrated significant improvements in model accuracy (i.e., positive transfer).\cite{Pilania-17-03,Smith-19-07,Egorova-20-10,Shui-22-10,Gardner-24-03,Cui-24-11,Alavi-25-01}
For example, Smith \textit{et al.} showed that pre-training on DFT energies and forces before fine-tuning on CC energies yields higher accuracy than training solely on CC energies.\cite{Smith-19-07}
Similarly, Alavi \textit{et al.} found that fine-tuning a DFT-pre-trained model with CC labels dramatically improves predictions of anharmonic vibrational frequencies.\cite{Alavi-25-01}
Positive transfer can also be obtained when using other low-fidelity data sources, such as empirical force fields,\cite{Shui-22-10} other MLFFs,\cite{Morrow-2022-IndirectLearning,Gardner-24-03,Gardner-2025-DistillationAtomistic} density functional tight binding methods,\cite{Cui-24-11} and different density functionals.\cite{Pilania-17-03,Egorova-20-10}
More recently, fine-tuning pre-trained foundation models on high-fidelity labels has also proven to be highly data-efficient.\cite{Yang-24-05}

\subsection{Joint Modelling}

Joint modelling approaches involve training a single model to predict properties from multiple labelling methods simultaneously.
A common strategy, particularly for MLFFs, is to use a shared model backbone (e.g., a GNN) that learns a common representation of each atom's chemical environment.
Multiple readout heads, typically linear layers or simple feed-forward networks, then map the shared representation to the specific property of interest for each fidelity (Fig.~\ref{fig:overview}b).\cite{Shoghi-23-10,Thaler-24-12,Chen-25-04,Huang-25-04,Batatia-2025-CrossLearning}
This architecture allows the model to learn from diverse datasets, potentially labelled with different quantum-chemical methods or covering distinct chemical domains.
For instance, Shoghi \textit{et al.} pre-trained a multi-headed model on several datasets, each labelled with a different density functional, demonstrating improved fine-tuning performance on downstream tasks.\cite{Shoghi-23-10}
Thaler \textit{et al.} and Chen \textit{et al.} also showed that multi-headed models can effectively learn from multiple fidelities, leading to error reductions, especially in low-data regimes.\cite{Thaler-24-12,Chen-25-04}
This approach is appealing for building foundation models as it naturally scales to incorporate an increasing number of datasets and fidelities.

Other joint modelling techniques have also been explored.
For example, some works have trained models on labels from multiple density functionals for predicting properties like band gaps or formation energies.\cite{Liu-24-04,Kim-24-09}
Hutchinson \textit{et al.} used multi-fidelity learning with random forests for band-gap prediction, combining DFT calculations and experimental data.\cite{Hutchinson-17-11}
More recently, Wood \textit{et al.} have proposed conditioning both the atomic embeddings and expert routing of their UMA model on the fidelity of the target label to train models that can generate predictions at multiple levels of theory using only a single readout head.\cite{Wood2026UMA}

\subsection{Other Multi-fidelity Approaches}

Beyond pre-training/fine-tuning and joint modelling, other multi-fidelity strategies have been explored.
A prominent architecture-agnostic alternative is $\Delta$-learning.\cite{Ramakrishnan-15-05, Hutchinson-17-11, Batra-19-07, Nandi-21-02, Wengert-21-04, Chen-23-05, Goodlett-23-07}
In this approach, a machine-learned model is trained to predict the residual between a low-fidelity base method and a high-fidelity target: the high-fidelity energy of a structure is approximated as the low-fidelity energy plus this learned correction term.
Because this correction is typically smoother and smaller in magnitude than the total energy, $\Delta$-models can be highly data-efficient and simple to implement.
However, the base method must be evaluated for every new structure at inference time, which limits the computational savings achievable during production simulations.

Other techniques applied to multi-fidelity learning in chemical applications include latent variable models and co-kriging, a statistically motivated approach based on Gaussian processes.\cite{Kennedy-00, Forrester-07-10, Pilania-17-03, Hutchinson-17-11, P.Greenman-22, Fare-22-12}
Generally, representation learning models, such as neural networks, tend to derive greater benefits from multi-fidelity training compared to traditional machine learning models like random forests.\cite{P.Greenman-22}

Specifically within the MLFF domain, Shiota \textit{et al.} recently introduced a multi-fidelity training strategy that employs a linear referencing scheme to align diverse labelling methods.\cite{Shiota-24-12}
This alignment enables merging multiple datasets to train a single model without architectural modifications.
The efficacy of this technique is inherently linked to the accuracy of the linear referencing scheme in aligning the different fidelities.

\section{Methodology}
\label{sec:methodology}

\subsection{Dataset}

\begin{figure}[ht]
    \centering
    \includegraphics[width=0.8\linewidth]{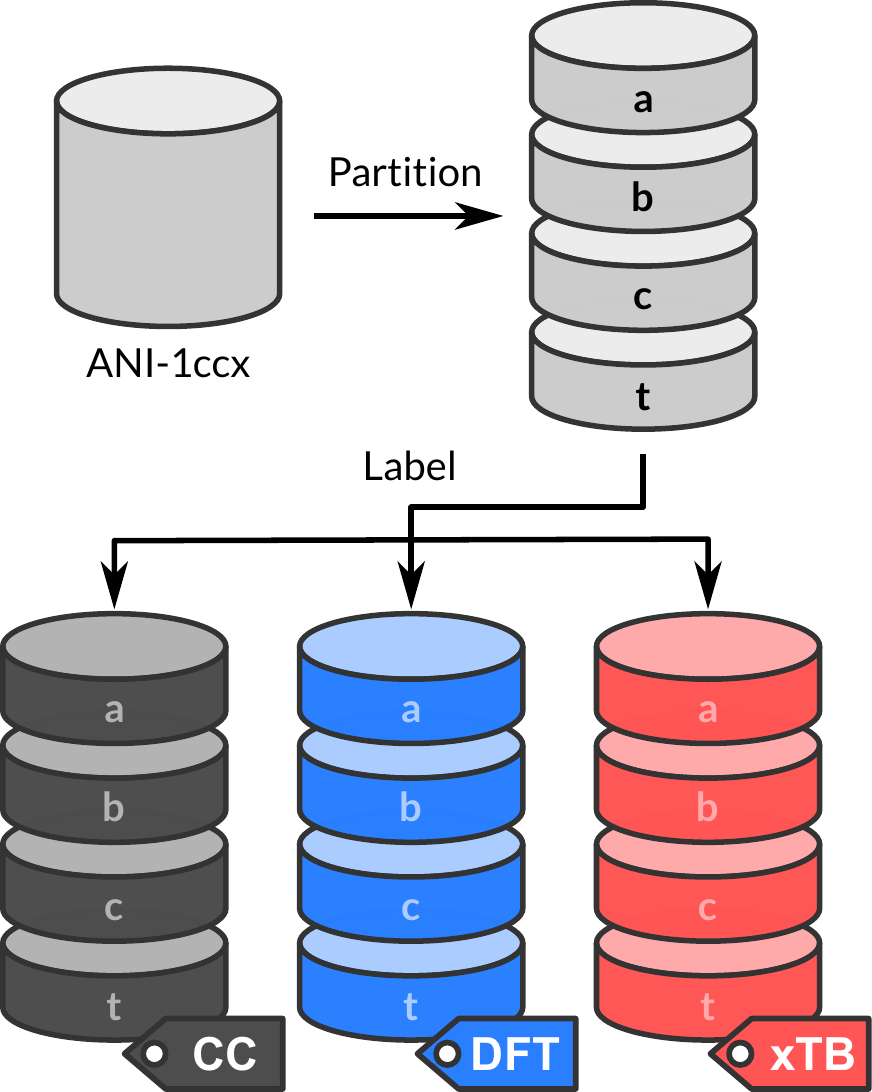}
    \caption{
        \label{fig:partitions}%
        Overview of the dataset partitioning and labelling scheme.
        We partition the dataset into four non-overlapping subsets termed \texttt{a}, \texttt{b}, \texttt{c}, and \texttt{t}.
        There are CC, DFT, and xTB labels for each structure in each subset.
        Unless otherwise stated, we test all models on the CC-labelled hold-out test set \texttt{t}.
        Subsets \texttt{a}, \texttt{c} and \texttt{t} contain \num{\approx 100}k structures each; the remaining \num{\approx 175}k structures were assigned to the subset \texttt{b}.
    }
\end{figure}

In this work, we used the ANI-1ccx dataset for training and evaluation.\cite{Smith-20-05,Smith-19-07}
This dataset comprises \num{\approx 500}k diverse conformations of small-to-medium-sized molecules.
Molecules are composed of carbon, hydrogen, nitrogen, and oxygen, and contain on average \num{14} atoms.
The dataset consists of \num{1910} unique chemical formulae.
Representative structures are shown in Fig.~\ref{fig:ani}.

Structures in this dataset are labelled with two quantum-chemical methods:
the $\omega$B97X approximate density functional with a \mbox{6-31G$^*$} basis set \cite{Chai-08-02,Ditchfield-71-01}
for energy and force labels,
and a composite extrapolation scheme based on the DLPNO-CCSD(T) method for an accurate energy label.\cite{Smith-20-05,Guo-18-01}
We refer to these labels as DFT and CC, respectively.

We ran GFN2-xTB calculations for each structure in the dataset to generate energy and force labels with a third, less accurate quantum-chemical method.\cite{Bannwarth-19-03, Bannwarth-21, Grimme-17-05, Pracht-19-06}
We refer to these labels as xTB.
Around \num{200} (\qty{< 0.05}{\percent}) of these calculations failed to converge; we discard these structures from our dataset, so that all remaining structures are labelled with all three methods (CC, DFT, and xTB).

We chose ANI-1ccx because it is, to the best of our knowledge, the only publicly available dataset that provides highly accurate CC labels as well as labels from another quantum-chemical method (here DFT) for the same set of structures.\cite{Smith-20-05,Smith-19-07}
Together with the xTB labels we computed, this yields three levels of accuracy, or fidelity, per structure, making the dataset uniquely suited to a systematic investigation of multi-fidelity training strategies.

For training and evaluation, we divided the structures into four partitions: \texttt{a}, \texttt{b}, \texttt{c}, and \texttt{t} (see Fig.~\ref{fig:partitions}).
We randomly assigned all structures with the same chemical formula to one of the four subsets;
this ensures that no structure appears in more than one subset and that structures from different subsets are sufficiently different.
Given the large number of unique chemical formulae in the dataset, this procedure produced four non-overlapping subsets of similar chemical diversity.
Subsets \texttt{a}, \texttt{c}, and \texttt{t} contain \num{\approx 100}k structures each;
the remaining \num{\approx 175}k structures were assigned to subset \texttt{b}.
Throughout, we use the notation (\textlangle{}label\textrangle{}; \textlangle{}subset\textrangle{}), for instance (CC; \texttt{a}) and (DFT; \texttt{t}), to refer to the various combinations of labelling methods and subsets.
Unless explicitly stated, we use energy and force labels when training on DFT and xTB, and evaluate models against the CC energy labels of the hold-out test split, \texttt{t}, since CC provides the most accurate labels and is the target fidelity throughout.

\subsection{Model Architectures}

Like most MLFFs, we model the energy $\hat{E}(s; \theta)$ of structure $s$ as a sum of $N_s$ (local) atomic contributions $\epsilon(a, s; \theta)$ with parameters $\theta$:
\begin{equation}
    \label{eq:decomposition}
    \hat{E}\left(s; \theta\right) = \sum_{a=1}^{N_s} \epsilon\left(a, s; \theta \right)
\end{equation}
To account for differences between labelling methods, we model the atomic energy $\epsilon$ from Eq.~\eqref{eq:decomposition} using a method-independent backbone $b$ together with a method-specific readout head $r$:
\begin{equation}
    \label{eq:backbone}
    \begin{aligned}
    \hat{E}^\mathcal{M} \big(s; \big(\theta_b, & \theta_r^\mathcal{M}, \mu^\mathcal{M} \big)\big) \\
    & = \sum_{a=1}^{N_s} \epsilon\left(a, s; \left(\theta_b, \theta_r^\mathcal{M}, \mu^\mathcal{M}\right)\right)  \\
    & = \sum_{a=1}^{N_s} r\left(b \left(a, s; \theta_b\right), \theta_r^\mathcal{M}\right) + \mu^\mathcal{M}_{Z_a}
    \end{aligned}
\end{equation}
The backbone contains the parameters $\theta_b$ that are independent of the method $\mathcal{M}$, whereas the readout head contains the parameters $\theta_r^\mathcal{M}$ that are specific to each $\mathcal{M}$.
$Z_a$ is the atomic number of atom $a$ and $\mu^\mathcal{M}$ is a method-dependent offset (cf.\ Table~\ref{tab:e0s} of the Appendix).

Unless otherwise stated, we use an adapted version of the MACE architecture with the following hyperparameters to model $\hat{E}^\mathcal{M}$:
2 message passing layers,
a radial cutoff of \qty{5}{\angstrom},
8~Bessel radial expansion functions,
256~hidden channels,
$\ell_\text{max} = 3$,
and a maximum correlation order of 3.\cite{Batatia-22}
The first readout layer is a linear map from the hidden channels to the output dimension, which is 1 for energy predictions.
The second readout layer is a multi-layer perceptron (MLP) with one hidden layer of 16 units and a SiLU activation function.
These settings result in a total of \num{\approx 3}$\,$M trainable parameters.

As part of our ablation studies, we also train models of the Allegro architecture.\cite{Musaelian-22-04}
We use a two-layer model with $\ell_\text{max}=2$,
a radial cutoff of \qty{5}{\angstrom},
8 radial basis functions,
and 64 and 256 channels for the equivariant and invariant features, respectively.
All readout heads are MLPs with one hidden layer of 128 units and a SiLU activation function.
These settings result in a total of \num{\approx 1}M trainable parameters.

\subsection{Loss Function}

For a given structure $s$, labelling method $\mathcal{M}$, and model with parameters $\theta$, we define the following loss:
\begin{equation}
    \begin{aligned}
        \ell^\mathcal{M}(s; \theta)
         & = \lambda_E^\mathcal{M} \left( \frac{\hat{E}^\mathcal{M}(s; \theta) - E^\mathcal{M}(s)}{N_s} \right)^2 \\
         & + \frac{\lambda_F^\mathcal{M}}{3 N_s} \sum_{\substack{1 \leq a \leq N_s                                \\ \alpha \in \{x, y, z\}}}
        \left( \hat{F}_{a, \alpha}^\mathcal{M}(s; \theta) - F_{a, \alpha}^\mathcal{M}(s) \right)^2
        ,
    \end{aligned}
    \label{eq:loss}
\end{equation}
where $\hat{E}^\mathcal{M}(s; \theta)$ and $E^\mathcal{M}(s)$ are the energies of structure $s$ given by the model with parameters $\theta$
and the labelling method $\mathcal{M}$, respectively.
Similarly, $\hat{F}^\mathcal{M}_{a, \alpha}(s; \theta)$ and $F_{a, \alpha}^\mathcal{M}(s)$ are the (conservative) force components of atom $a$ along the axis $\alpha$ for $s$ given by the model and the method $\mathcal{M}$, respectively.
$\lambda_E^\mathcal{M}$ and $\lambda_F^\mathcal{M}$ are weighting constants for the per-atom energy and force components of the loss, respectively.
If forces are available, we choose $\lambda_F^\mathcal{M} = 7$ as a trade-off between the two loss components in our experiments.
If forces are unavailable, we formally set $\lambda_F^\mathcal{M} = 0$; in practice, this loss component is not evaluated in that case.
We set $\lambda_E^\text{DFT} = \lambda_E^\text{xTB} = 1$, but $\lambda_E^\text{CC} = 100$.
We choose a higher weight for $\lambda_E^\text{CC}$ because CC forces are unavailable.
This way the total losses of all $\mathcal{M}$ are of similar magnitude.

In this study, we train models on different partitions $d$ of the ANI-1ccx dataset (cf.\ Fig.~\ref{fig:partitions}) labelled with different methods $\mathcal{M}$:
$\mathcal{D} \subseteq \left\{
    (\mathcal{M}, d) \, | \,
    \mathcal{M} \in \left\{ \text{CC}, \text{DFT}, \text{xTB} \right\},
    d \in \left\{ \texttt{a},\texttt{b}, \texttt{c}, \texttt{t} \right\}
    \right\}$.
The loss with which we optimise the model parameters $\theta$ reads:
\begin{equation}
    \mathcal{L}(\mathcal{D}, \theta) =
    \sum_{(\mathcal{M}, d) \in \mathcal{D}} \sum_{s \in d} \ell^{\mathcal{M}}(s; \theta) \, p(\mathcal{M}, d, s),
    \label{eq:total_loss}
\end{equation}
where $p(\mathcal{M}, d, s) = p(s | \mathcal{M}, d) p(\mathcal{M}, d)$ is the probability of sampling structure $s$ from partition $d$ labelled with method $\mathcal{M}$ during training, and $\ell^\mathcal{M}(s; \theta)$ is the loss defined in Eq.~\eqref{eq:loss}.
Unless stated otherwise, we uniformly sample $(\mathcal{M}, d)$ from an experiment-specific $\mathcal{D}$.
Then the probability of sampling structure $s$ given $\mathcal{M}$ and $d$, $p(s | \mathcal{M}, d)$, is always set to $1 / |d|$ if $s \in d$ and $0$ otherwise, where $|d|$ is the number of elements in $d$.
We note that an alternative approach to control the relative importance of different fidelities is to vary their respective loss component weights.
These two loss formulations share the same expectation, but differ in variance.

\subsection{Training}

We use the \verb|Adam| optimiser as implemented in \verb|PyTorch|.\cite{Kingma-17-01, Paszke-19-12}
Unless otherwise stated, the initial learning rate is $10^{-2}$ for MACE and $10^{-3}$ for Allegro.
We apply a learning rate schedule that reduces the rate by a factor of 0.8 whenever the validation loss has not improved for 32k steps, and terminate training after 100k steps without improvement.

For all model architectures and datasets considered in this work, we found that pre-training to completion (i.e., until the validation loss stops improving) on the pre-training task led to the best performance upon fine-tuning.

We investigate several fine-tuning strategies below;
in all cases, we reinitialise the optimiser's parameters before fine-tuning and use an initial learning rate of $10^{-3}$ (see Fig.~\ref{fig:repr-lr} in the Appendix for motivation).

All experiments are repeated with three independent random seeds; reported values are means unless otherwise stated.
Where error bars are not shown, inter-seed variance is smaller than the marker size.

We train models on four NVIDIA V100 GPUs in parallel.
We employ the ``Worst Fit'' online bin-packing algorithm with 8 bins to create batches of structures with the same labelling method.\cite{johnson1973near}
Each batch contains up to 256 atoms, which allows us to maximise GPU utilisation without memory overflows due to the wide distribution of structure sizes in the ANI-1ccx dataset.
In our distributed setting, the batches produced by the bin packer are assigned to the GPUs in round robin order to ensure each data point is seen exactly once per epoch.

\begin{figure*}[!ht]
    \centering
    \includegraphics{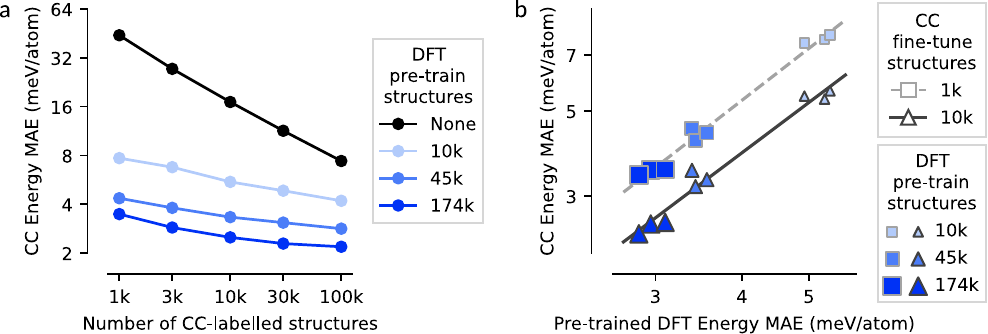}
    \caption{\label{fig:amount}%
        Effect of quantity of pre-training data on accuracy of fine-tuned model.
        \textbf{(a)} Mean absolute error (MAE) in CC energy predictions versus the number of CC-labelled fine-tuning structures.
        Coloured lines connect points corresponding to the same amount of pre-training data: black for no pre-training (i.e., direct CC training), and increasingly dark shades of blue for increasing amounts of DFT-labelled structures used during pre-training.
        \textbf{(b)} Relationship between the energy MAE of the pre-trained model on the DFT-labelled validation set and the final fine-tuned model's energy MAE on the CC-labelled test set.
        Each point corresponds to a single run, with larger markers and darker colours indicating more pre-training data.
        Separate log-log linear best fits are shown for models fine-tuned on 1k (squares) and 10k (triangles) CC data.
    }
\end{figure*}

\section{Pre-training and Fine-tuning Ablation Studies}
\label{sec:pt-ft}

We begin by extending the pre-train/fine-tune experiment of Smith \textit{et al.}, \cite{Smith-19-07} using the MACE architecture instead of ANI, and ablating both the amount of DFT-labelled pre-training data (taken from split \texttt{b}) and of CC-labelled fine-tuning data (taken from split \texttt{a}).
We benchmark against models directly trained on the CC-labelled data (i.e., without pre-training, black line), and compare the accuracies of all final models on the hold-out CC-labelled test set \texttt{t} in Fig.~\ref{fig:amount}a.

We make several observations from these results:
first, even small amounts of pre-training data (10k DFT labels) lead to greatly reduced CC test-set errors compared to direct training.
This positive transfer is strongest in the low-CC-data regime (85\% reduction in error for 1k CC labels), but remains significant even in the high-CC-data regime (50\% reduction in error for 100k CC labels).

Second, pre-training on more data further improves accuracy after fine-tuning:
compared to direct training, pre-training on 174k DFT-labelled structures reduces CC test-set error by 92\% when fine-tuning on 1k CC labels, and by 70\% for 100k CC labels.

Third, fine-tuned accuracy can be improved, and overall labelling cost reduced, by simultaneously increasing the amount of pre-training data and decreasing the amount of fine-tuning data.
For example, pre-training on 45k DFT and fine-tuning on 1k CC requires fewer labels and generates a more accurate model than pre-training on 10k DFT and fine-tuning on 100k CC.
We provide a full cost-error analysis in Fig.~\ref{fig:cost-curve-xtb-dft} of the Appendix.

Finally, we briefly compare our results to those obtained by Smith \textit{et al.}\cite{Smith-19-07} using the ANI architecture in Table~\ref{tab:ani-vs-mace}.
While an exact comparison is not possible due to differences in the amounts and partitioning of the training data, we find that MACE demonstrates much higher degrees of positive transfer.

We now investigate the mechanism by which pre-training on more DFT-labelled data leads to stronger positive transfer.
One explanation could be that pre-training on more data leads to models with lower initial error on the CC labels even before any fine-tuning.
We find that this is not the case: all pre-trained models, regardless of the amount of pre-training data, achieve similar accuracy on the CC-labelled test set, with energy MAEs of \qty{18 \pm 1}{\mevpatom} when accounting for the different method-dependent offsets, $\mu^\mathcal{M}$ (cf.\ Table~\ref{tab:e0s} of the Appendix).

Instead, we find that pre-training on more data leads to models that are better at the \textit{pre-training task}, and hence to models with improved internal representations of local chemical environments.
These improved representations then translate into lower error on the CC test set after fine-tuning.
To demonstrate this mechanism, in Fig.~\ref{fig:amount}b, we plot the pre-trained model's accuracy on the pre-training task (measured as the energy MAE on the DFT-labelled validation set) and the fine-tuned model's energy MAE on the CC-labelled test set.

For a given amount of fine-tuning data, we observe a strong log-log linear relationship between pre-trained ($x$) and fine-tuned ($y$) model error:
\begin{equation}
    \log(y) = m \log(x) + c
    \label{eq:log_log_linear_power_law}
\end{equation}
The gradient, $m$, appears to be independent of the amount of fine-tuning data, and quantifies how strongly improvements in the pre-training task translate into improvements in the fine-tuning task.
Here, we see a gradient of 1.35, indicating that a 50\% reduction in pre-training error leads to a $1 - 0.5^{1.35} \approx 60$\% reduction in fine-tuning error.

Below, we perform ablation studies to investigate the effect of various factors on the degree of positive transfer, and to test whether the log-log linear relationship between pre-training and fine-tuning performance observed in Fig.~\ref{fig:amount}b and presented in Eq.~\eqref{eq:log_log_linear_power_law} is robust to these factors.
Concretely, we now seek to isolate the effects of:

{
\raggedright
\begin{itemize}\setlength{\itemsep}{0pt}\setlength{\parskip}{0pt}
    \item the alignment between the low-fidelity and high-fidelity labels,
    \item the size of the model,
    \item the model architecture, and
    \item the types of labels (i.e., energies and/or forces) used during pre-training.
\end{itemize}
}
In the Appendix, we also investigate the importance of model capacity and learning rate during fine-tuning.

\subsection{Labelling Method Used For Pre-training}

\begin{figure*}[!h]
    \centering
    \includegraphics{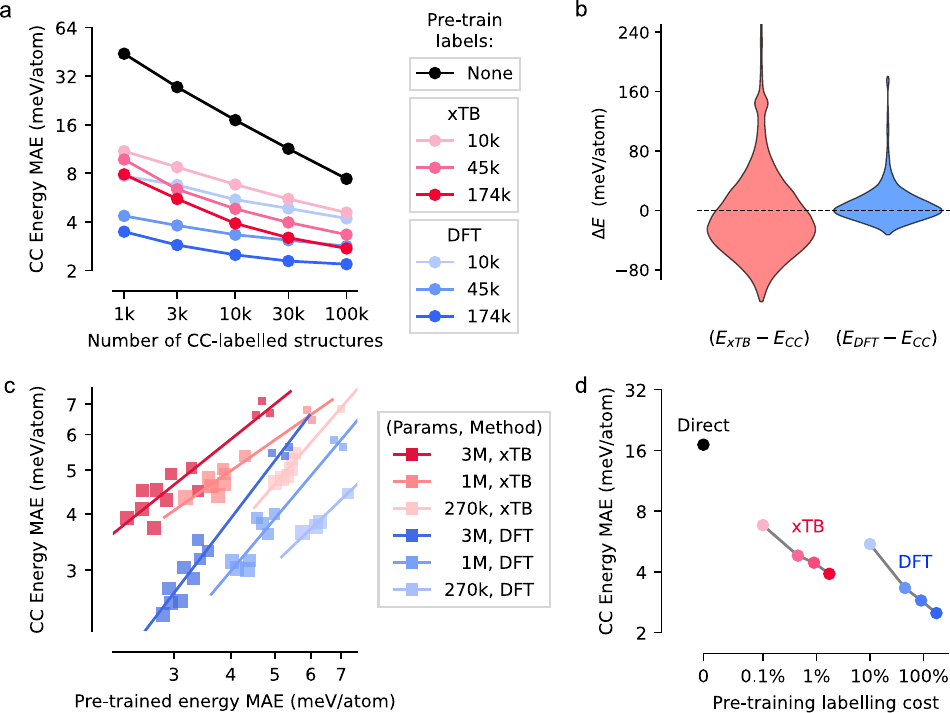}
    \caption{\label{fig:dft-vs-xtb}%
        Effect of pre-training labels on accuracy after fine-tuning.
        \textbf{(a)} CC test set MAE versus number of CC-labelled fine-tuning structures.
        Line colour indicates the amount and source of pre-training data: black for no pre-training; increasingly dark shades of blue and red for more DFT- and xTB-labelled structures from split \texttt{b}, respectively.
        \textbf{(b)}
        Distribution of energy differences between DFT/xTB and CC labels after accounting for energy offsets $\mu^{\mathcal{M}}$.
        \textbf{(c)}
        Pre-trained model error on the relevant DFT/xTB validation set versus fine-tuned model error on the CC test set for models of varying size, pre-trained on various amounts of (DFT/xTB; \texttt{b}) and fine-tuned on 10k (CC; \texttt{a}).
        Each point is a single run, with larger markers corresponding to more pre-training data.
        Log-log lines of best fit shown per model size and labelling method.
        \textbf{(d)}
        CC test MAE versus additional cost of generating pre-training labels.
        The cost is relative to that of the 10k fine-tuning labels, assuming xTB:CC and DFT:CC cost ratios of 1:1000 and 1:10, respectively.
    }
\end{figure*}

We investigate the effect of pre-training models on a more approximate method than DFT.
To this end, we pre-train a series of models on varying amounts of structures from (xTB; \texttt{b}) before fine-tuning on increasing amounts of (CC; \texttt{a}).
The results of this experiment are shown in Fig.~\ref{fig:dft-vs-xtb}a, and compared to the analogous experiment we reported in Fig.~\ref{fig:amount}a using DFT labels for pre-training.

For all amounts of pre-training and fine-tuning data, we find that pre-training on DFT labels yields greater positive transfer than pre-training on the same structures labelled by xTB.
This difference is most pronounced when fine-tuning on small amounts of CC-labelled data (1--3k structures).
We posit that this difference is due to the greater alignment between the DFT and CC energies, compared to the xTB and CC energies:
accounting for method-specific energy offsets $\mu^{\mathcal{M}}$, the mean absolute difference between CC and DFT energies and between CC and xTB energies is
\num{18.3} and \qty{47.3}{\mevpatom}, respectively (Fig.~\ref{fig:dft-vs-xtb}b).
The two residuals are, moreover, only moderately correlated (Pearson $r=0.52$), suggesting that there is no simple linear relationship between the xTB and DFT errors with respect to CC.

To further understand the impact of the pre-training labelling method, we plot the relationship between the pre-training validation error and the fine-tuning test error for models of varying size, pre-trained on various amounts of (DFT/xTB; \texttt{b}) and fine-tuned on 10k (CC; \texttt{a}) in Fig.~\ref{fig:dft-vs-xtb}c.
We make several observations from this analysis:
first, in all cases, we continue to observe a strong log-log linear relationship between pre-trained and fine-tuned error, such that a consistent improvement in the pre-training task translates into a consistent improvement in the fine-tuning task.

Second, comparing xTB to DFT pre-training, we find that the pre-factors of the xTB relationships are higher than those for DFT, and that the xTB gradients are shallower.
These indicate respectively that (i) for a given pre-training error, xTB pre-training leads to worse fine-tuning performance than DFT pre-training, and (ii) improvements in the xTB pre-training task translate into smaller improvements in the fine-tuning task than improvements in the DFT pre-training task.
We attribute these differences again to the worse alignment of xTB labels with the CC labels, compared to DFT labels.

Third, we find that larger models (as measured by parameter count) have steeper gradients than smaller models, i.e., that larger models are better able to translate improvements in the pre-training task into improvements in the fine-tuning task.
Interestingly, smaller models have lower pre-factors than larger models, i.e., for a given pre-training error, smaller models achieve better fine-tuning performance than larger models.
We propose that this is due to some combination of the following two factors:
(i) smaller models have to be pre-trained on more structures to achieve the same accuracy on the pre-training task as larger models, and
(ii) smaller models are less prone to overfitting on the energy labels during fine-tuning, leading to better generalisation on the test set.

Finally, while pre-training on xTB labels leads to worse fine-tuned performance than DFT labels for a given amount of pre-training data, xTB is computationally more affordable than DFT.
Therefore, if one considers the computational cost of generating the pre-training labels, xTB pre-training can be a sensible choice for low pre-training budgets (see Fig.~\ref{fig:dft-vs-xtb}d, and also Fig.~\ref{fig:cost-curve-xtb-dft} of the Appendix for a more detailed cost-error analysis).

\clearpage

\begin{figure*}[!ht]
    \centering
    \includegraphics{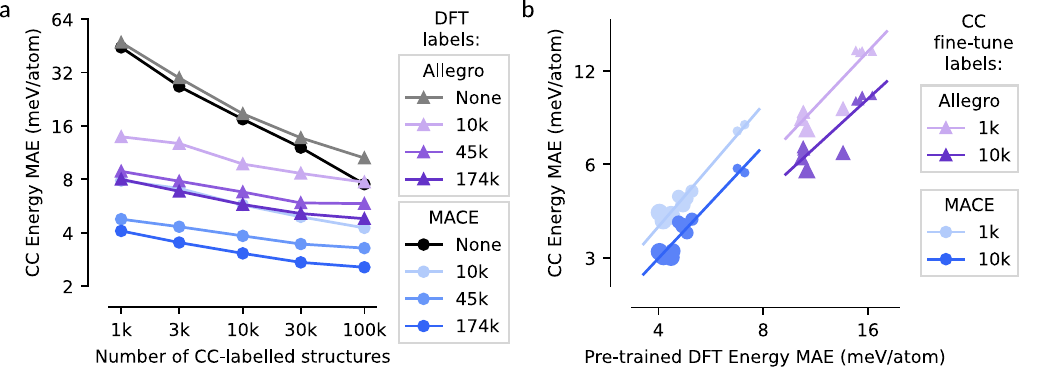}
    \caption{\label{fig:allegro-vs-mace}%
        Effect of model architecture on the benefits of pre-training.
        \textbf{(a)}
        CC test MAE for MACE and Allegro models pre-trained on (DFT; \texttt{b}) and fine-tuned on (CC; \texttt{a}), varying the amounts of pre-training and fine-tuning data.
        Black lines show direct training on (CC; \texttt{a}) for reference.
        \textbf{(b)}
        Pre-trained error on the DFT validation set versus fine-tuned error on the CC test set for both architectures, each fine-tuned on either 10k or 100k CC structures.
        Each point corresponds to an independent training run, with larger markers corresponding to more pre-training data.
        Log-log lines of best fit are shown for each architecture and amount of fine-tuning data.
        }
\end{figure*}

\clearpage

\subsection{Model Architecture}

To test whether the pre-training mechanism identified above depends on the choice of model architecture, we repeat the experiment from Fig.~\ref{fig:amount} using Allegro in place of MACE, with hyperparameters chosen so that both models have approximately 1M trainable parameters (Fig.~\ref{fig:allegro-vs-mace}).

In a direct training setting, both architectures perform similarly when trained on small amounts of CC data,
but Allegro's accuracy plateaus as the training set grows, while MACE continues to improve.
Pre-training on DFT data benefits both architectures, but the gains are substantially larger for MACE than for Allegro (Fig.~\ref{fig:allegro-vs-mace}a).
This difference is explained by Allegro's worse performance on the pre-training task: as shown in Fig.~\ref{fig:allegro-vs-mace}b, Allegro achieves a higher energy MAE on the DFT validation set than MACE.
Nevertheless, when calibrating pre-training accuracy against fine-tuning accuracy, both architectures follow the same log-log linear relationship, with similar gradients for both architectures and amounts of fine-tuning data: combined with Fig.~\ref{fig:dft-vs-xtb}c, this suggests that the efficacy of positive transfer is determined primarily by the size of the model, and the kind of pre-training label, rather than any architecture-specific property.

\subsection{Training on Forces}

\begin{figure}[ht]
    \centering
    \includegraphics[width=\linewidth]{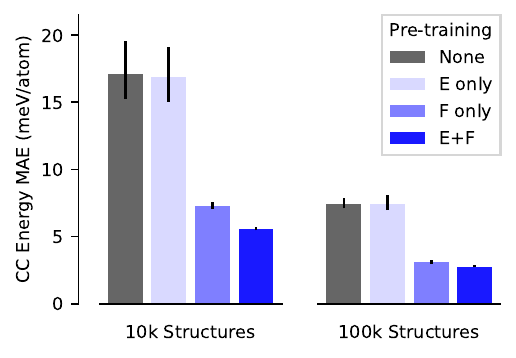}
    \caption{\label{fig:ablation-forces}%
        Effect of energy and force labels during DFT pre-training.
        Bars show the CC test set error; error bars span the min--max range.
        Left and right panels correspond to 10k and 100k training structures, respectively.
        Grey bars: direct training on (CC; \texttt{a}).
        Blue bars: pre-training on (DFT; \texttt{a}) followed by fine-tuning on (CC; \texttt{a}).
        Shading indicates the label type used during pre-training: light for energies only, medium for forces only, dark for both.
        See Fig.~\ref{fig:ablation-forces-xtb} in the Appendix for an analogous experiment with xTB labels.
    }
\end{figure}

The pre-training experiments above used both DFT energy and force labels.
To isolate the contribution of each label type, we pre-train models on the same structures as we fine-tune on, and ablate the use of energy and force labels during pre-training (Fig.~\ref{fig:ablation-forces}).

Pre-training on DFT energies alone, and on the \textit{same} structures as used for fine-tuning, provides no improvement over direct CC training (light bars in Fig.~\ref{fig:ablation-forces}).
We therefore find that a second set of energy labels for identical structures confers no benefit, for example through implicit regularisation.
Pre-training on forces alone, by contrast, already substantially improves CC energy predictions (medium bars).
This result is consistent with the well-established observation that force labels, which contribute $3N$ components per structure versus a single energy, drive model training; our experiment confirms that the same holds in the pre-train/fine-tune setting.

\begin{figure*}[ht]
    \centering
    \includegraphics{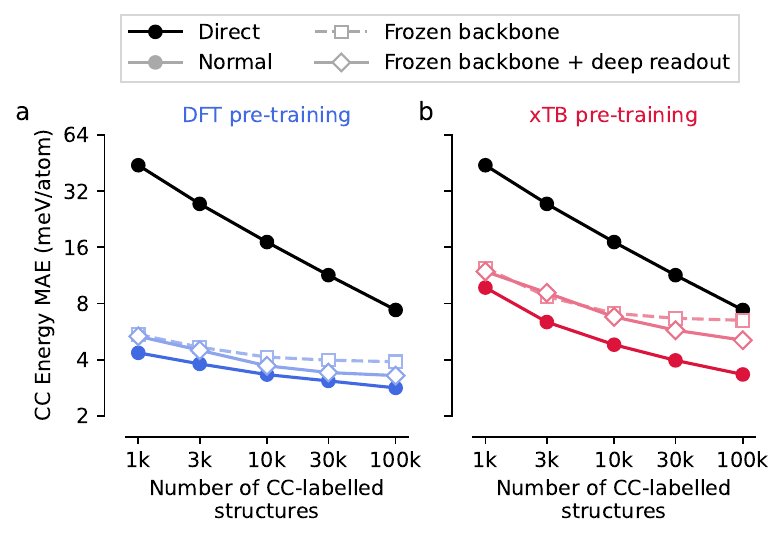}
    \caption{\label{fig:repr:backbone}%
        Effect of backbone freezing during fine-tuning on CC test set error (y-axis).
        Models are pre-trained on 45k structures from (DFT; \texttt{b}) (a) and (xTB; \texttt{b}) (b) and then fine-tuned on increasing amounts of (CC; \texttt{a}) (x-axis).
        Coloured solid lines show full model fine-tuning.
        Dashed lines ($\square$) show backbone-frozen fine-tuning with only the readout heads updated.
        Solid lines ($\scalebox{1.2}{$\diamond$}$) show backbone-frozen fine-tuning with an enlarged final readout head (three hidden layers, each 128 channels wide, leading to 62k additional parameters).
        The black line ($\scalebox{1.2}{$\bullet$}$) shows direct training on (CC; \texttt{a}) without pre-training.
    }
\end{figure*}

Including both energies and forces during pre-training yields the largest gains, reducing the error by up to 65\% (dark bars).
The additional benefit of energy labels, absent when used without forces, suggests that energies and forces provide complementary supervision: forces constrain the local curvature of the potential energy surface, while energies anchor its global scale.
This complementarity is also reflected in training robustness: the error bars in Fig.~\ref{fig:ablation-forces} reveal that inter-seed variance is much larger when pre-training on energies only, underscoring the role of force labels in stabilising the learned representations.
All these observations hold for both 10k and 100k training structures and when replacing DFT with xTB (Fig.~\ref{fig:ablation-forces-xtb}).

{\raggedright
\section{Method-Specificity of Pre-trained Representations}
\label{sec:method-specificity}
}

The preceding sections established that pre-training on DFT or xTB labels improves performance after fine-tuning on CC labels, owing to improved internal representations of local chemical environments learned during pre-training.
We now investigate whether these representations are specific to the pre-training quantum-chemical method or transferable across methods.

As illustrated in Fig.~\ref{fig:overview} and formulated in Eq.~\eqref{eq:backbone}, we decompose the model into a backbone and a readout head.
The backbone maps each atom's local chemical environment to a latent representation; the readout head regresses this representation to predict the atom's energy contribution.

To probe the label specificity of the backbone, we ran new experiments in which we froze the backbone parameters during fine-tuning and optimised only the readout heads, $\theta_r^\mathcal{M}$.
If the backbone representations are transferable across labelling methods, performance should not degrade; if they are label-specific, performance should suffer relative to full fine-tuning.
Results for both DFT and xTB are shown in Fig.~\ref{fig:repr:backbone}, where we pre-train on 45k structures from split \texttt{b} and fine-tune separately on increasing amounts of (CC; \texttt{a}).

For both DFT and xTB pre-training, freezing the backbone during fine-tuning still leads to positive transfer, confirming that the pre-trained representations encode information useful for modelling the CC potential energy surface.
However, in all cases, backbone freezing leads to worse performance than full fine-tuning.
Furthermore, backbone freezing is more detrimental after xTB pre-training than after DFT pre-training, consistent with the weaker alignment between the xTB and CC potential energy surfaces (Fig.~\ref{fig:dft-vs-xtb}), and indicating that xTB pre-training leads to representations that require greater adaptation to model the CC labels.
Enlarging the readout head partially closes the gap between backbone-frozen and fully fine-tuned models, but the accuracy of backbone-frozen models plateaus with increasing fine-tuning data, while full fine-tuning continues to improve.

These observations indicate that pre-trained representations are indeed label-specific and require adaptation to optimally model the target CC labels, consistent with recent work.\cite{Radova-25-02}

\section{Learning Universal Representations via Multi-headed Training}
\label{sec:multi-headed}

\begin{figure*}
    \centering
    \includegraphics{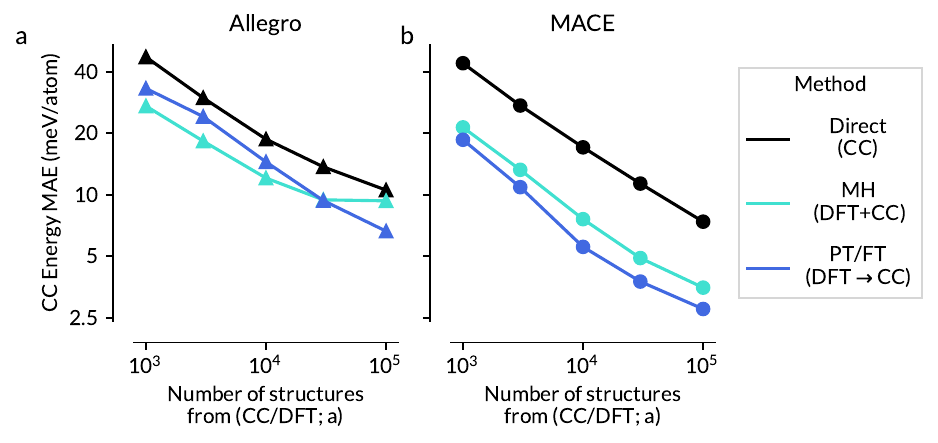}
    \caption{\label{fig:mh-mace-vs-allegro}%
        Comparison of CC test set error (y-axis) for direct training (black), multi-headed training (turquoise), and pre-train/fine-tuning (blue) for Allegro (a) and MACE (b).
        All models are trained on $N$ structures from split \texttt{a} (x-axis).
        Direct training uses (CC; \texttt{a}); multi-headed training uses (CC; \texttt{a}) and (DFT; \texttt{a}) simultaneously; pre-train/fine-tune involves pre-training on (DFT; \texttt{a}) followed by fine-tuning on (CC; \texttt{a}).
    }
\end{figure*}

The previous section established that pre-trained representations are method-specific and require adaptation during fine-tuning.
We now investigate whether multi-headed training (Fig.~\ref{fig:overview}b) can learn method-independent backbone representations that are useful across labelling methods.

We train multi-headed models on structures from a single split (\texttt{a}), providing both CC energy labels and DFT energy and force labels for each structure.
The model is thus encouraged to learn an internal representation agnostic to the labelling method, encoding method-specific information within the respective readout heads.
Results for both Allegro and MACE are shown in Fig.~\ref{fig:mh-mace-vs-allegro}; we discuss each architecture separately, as they exhibit qualitatively different behaviour.

For Allegro, multi-headed training consistently improves performance on the CC test set relative to direct training.
In the low-data regime, the multi-headed model outperforms the pre-train/fine-tune approach; however, as training data increases, multi-headed performance plateaus while pre-train/fine-tune performance continues to improve.
This plateauing suggests a capacity limitation in the 1M-parameter Allegro model.
When subsequently fine-tuned on (CC; \texttt{a}) alone, we find that the multi-headed models recover the performance of the pre-train/fine-tune models.

For MACE, multi-headed training yields a \qty{55}{\percent} reduction in energy MAE on the CC test set relative to direct training (Fig.~\ref{fig:mh-mace-vs-allegro}), demonstrating that the shared backbone successfully learns method-independent representations that benefit the CC task.
However, the pre-train/fine-tune approach achieves a larger reduction of \qty{65}{\percent}, indicating that the shared representations of the multi-headed model, while effective, are not optimal for any single fidelity.

Table~\ref{tab:training-error} quantifies this compromise on the training set.
A single-headed MACE model trained on 100k (CC; \texttt{a}) achieves a training MAE of \qty{0.24}{\mevpatom}, whereas a multi-headed model trained concurrently on 100k (CC; \texttt{a}) and 100k (DFT; \texttt{a}) yields a CC energy MAE of \qty{1.06}{\mevpatom} on the same structures.
Therefore, the multi-headed model's shared backbone must compromise between the two quantum-chemical methods, leading to a more than fourfold increase in training error on the CC labels.

\begin{table}[ht]
    \centering
    \caption{\label{tab:training-error}%
        Training MAEs on subset \texttt{a} for MACE models trained on 100k structures with different label combinations.
        Errors are reported for CC energies (E), DFT energies, and DFT forces (F) where applicable, in \unit{\mevpatom} for energies and \unit{\mevpang} for forces.
    }
    \begin{tabular}{@{}l|ccc@{}}
        \toprule
        Training labels & CC (E) & DFT (E) & DFT (F) \\
        \midrule
        CC only         & 0.24   & -       & -       \\
        CC + DFT        & 1.06   & 2.76    & 22.2    \\
        DFT only        & -      & 2.60    & 19.7    \\
        \bottomrule
    \end{tabular}
\end{table}

In summary, multi-headed training successfully learns method-independent backbone representations, providing consistent positive transfer from auxiliary DFT labels to the target CC task for both architectures.
The degree of transfer depends on the dataset and architecture, with Allegro benefiting more than MACE from multi-headed training.
These shared representations are, however, not optimal for any single fidelity: in the large-data regime, fine-tuning, which allows the backbone to specialise, achieves higher accuracy.

\begin{figure}[!ht]
    \centering
    \includegraphics{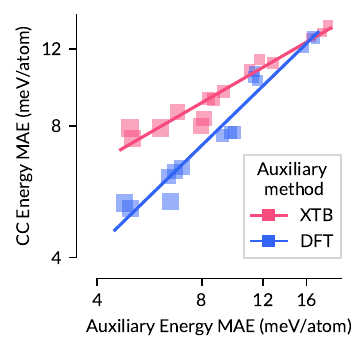}
    \caption{\label{fig:mh-val-aux-vs-cc}%
        Relationship between auxiliary and CC energy errors for multi-headed MACE models.
        Each marker corresponds to a unique model trained on 10k (CC; \texttt{a}) together with DFT-labelled (blue) or xTB-labelled (red) structures from a separate data split.
        Larger markers correspond to more auxiliary training data.
        The energy MAE on the DFT/xTB test set is plotted on the $x$-axis; the MAE on the CC test set on the $y$-axis.
        Log-log lines of best fit are shown for DFT and xTB separately.
    }
\end{figure}

{\raggedright
\subsection{Measuring Transfer Between Heads}
}

Having established that multi-headed training provides positive transfer from auxiliary labels to the target task, we now quantify this effect by examining how the accuracy of the auxiliary head relates to that of the target's head across models.
Fig.~\ref{fig:mh-val-aux-vs-cc} shows this relationship for multi-headed MACE models trained on 10k (CC; \texttt{a}) with increasing amounts of DFT- or xTB-labelled structures from a separate split.
For both auxiliary labelling methods, a log-log linear relationship again emerges: as the auxiliary head's error decreases, the CC head's error decreases proportionally.
We compare the gradients of these relationships to those observed in the pre-train/fine-tune setting (Fig.~\ref{fig:dft-vs-xtb}c) in Table~\ref{tab:gradients}.

\begin{table}[h]
    \centering
    \caption{\label{tab:gradients}%
        Gradients of the log-log relationships between auxiliary and CC error for models trained on 10k (CC; \texttt{a}) for pre-train/fine-tune and multi-headed settings.
    }
\begin{tabular}{l|cc}
\toprule

& \shortstack{pre-train/\\fine-tune} & multi-head  \\

 \midrule

xTB & 0.81 & 0.46 \\
DFT & 1.35 & 0.77 \\

\bottomrule
\end{tabular}

\end{table}

As before, we see a steeper gradient for DFT than for xTB, indicating that a given relative improvement in the auxiliary head's accuracy translates into a greater relative improvement in the CC head's accuracy when using DFT labels than when using xTB labels.

Comparing the pre-train/fine-tune and multi-headed settings is also instructive:
we find shallower gradients (by a factor of 0.6) in the multi-headed setting, demonstrating that the efficiency of positive transfer from the auxiliary to the target task is lower when the backbone must serve both heads simultaneously.

\subsection{Effect of Relative Sampling Rates}

Multi-headed training requires sampling from multiple datasets.
In this ablation study, we investigate how preferentially sampling from one dataset over another affects the accuracy of each readout head.
To this end, we set a specific $p(\mathcal{M}, d)$ in Eq.~\eqref{eq:total_loss}:
\begin{align}
    p(\text{DFT}, \texttt{b}) & = \frac{w_\text{DFT}}{w_\text{DFT} + 1} \\
    p(\text{CC}, \texttt{a})  & = 1 - p(\text{DFT}, \texttt{b})
\end{align}
with $p(\mathcal{M}, d) = 0$ for all other combinations.
The hyperparameter $w_\text{DFT} \in \mathbb{R}^+$ controls the relative sampling rate of DFT-labelled structures.

\begin{figure}[!h]
    \centering
    \includegraphics{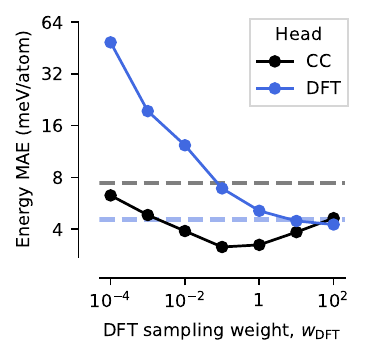}
    \caption{\label{fig:mh-sampling-rate}%
        Effect of the DFT sampling weight $w_\text{DFT}$ on the energy MAE of the CC and DFT heads after multi-headed training on (CC; \texttt{a}) and (DFT; \texttt{b}).
        Dashed lines indicate the MAE of single-headed models trained on CC (black) or DFT (blue) data alone.
    }
\end{figure}

We train multi-headed models on all of (CC; \texttt{a}) and (DFT; \texttt{b}), varying $w_\text{DFT}$ from $10^{-4}$ to $10^2$, and measure the energy MAE of each head on the corresponding test set (Fig.~\ref{fig:mh-sampling-rate}).
When $w_\text{DFT}$ is low, the CC head approaches the accuracy of a single-headed CC model (black dashed line) while the DFT head performs poorly.
At the other, DFT-dominated, extreme ($w_\text{DFT} = 10^2$, i.e., 99\% DFT batches), the CC head still outperforms a model trained on CC data alone, demonstrating that positive transfer persists across a wide range of sampling rates.

An optimal range around $w_\text{DFT} \approx 1$ maximises performance on the CC task while maintaining acceptable accuracy on the DFT task.
This region is broad, suggesting that multi-headed training is robust to the precise choice of sampling weight.
More generally, the sampling rate provides a practical lever for controlling the relative importance of each dataset during training, which may be desirable when one dataset is more relevant to the target task or contains more accurate labels.
A similar trend is observed for xTB auxiliary data (see Fig.~\ref{fig:mh-sampling-rate-xtb} in the Appendix).

\begin{figure*}[htb]
    \centering
    \includegraphics{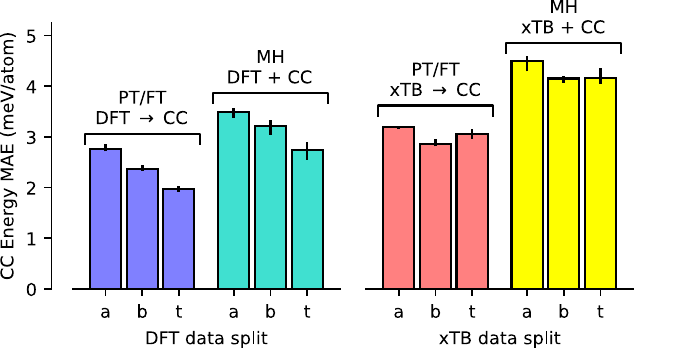}
    \caption{\label{fig:ablation-variety}%
        Effect of structural overlap between target and auxiliary datasets on CC test set performance.
        All models are trained on 100k (DFT/xTB; \texttt{a/b/t}) and 100k (CC; \texttt{a}), using either pre-train/fine-tune (PT/FT) or multi-headed (MH) training.
        Error bars show the min--max range.
    }
\end{figure*}
{\raggedright
\subsection{Structural Overlap Between Datasets}
}

We now investigate how structural overlap between the target and auxiliary datasets affects final model performance.
To do this, we train using 100k CC labels from split \texttt{a}, together with 100k auxiliary labels from one of splits \texttt{a}, \texttt{b}, or \texttt{t}.
We present final performance on the CC-labelled test set \texttt{t} under both pre-train/fine-tune and multi-headed training, for both DFT and xTB auxiliary labels, in Fig.~\ref{fig:ablation-variety}.

Consider first the pre-train/fine-tune setting using DFT auxiliary labels (blue bars in Fig.~\ref{fig:ablation-variety}).
Pre-training on DFT-labelled structures from a different split (\texttt{b}) compared to the same split (\texttt{a}) as the CC-labelled data improves final performance on the CC test set (\texttt{t}) by 15\%.
We explain this by noting that models pre-trained on \texttt{b} have seen twice as many unique structures during the entire training process, allowing them to learn more general representations of atomic structure.

Testing the limits of this approach, we find that pre-training on the DFT labels of the test-set structures themselves leads to a further modest improvement of 15\% in final performance:
during pre-training, the model learns representations that are explicitly tailored to the test-set structures.
Fine-tuning then adapts these representations to the new structures (from split \texttt{a}) and task (CC prediction), but retains some useful information about the test-set geometries from pre-training.
We note, however, that this approach does not reduce the error to zero: training on low-fidelity labels in the target domain does not lead to perfect generalisation to high-fidelity labels.

We find that these trends are qualitatively identical under multi-headed training when using DFT (turquoise bars in Fig.~\ref{fig:ablation-variety}): seeing more unique structures (by training on DFT labels from a different split) reduces error by improving the shared backbone representations, and seeing DFT-labelled test-set structures themselves leads to a further modest improvement in performance on the CC test set.

Finally, consider the use of xTB labels (orange and yellow bars in Fig.~\ref{fig:ablation-variety}).
We again find that sampling xTB-labelled structures from a different split (\texttt{b}) compared to the CC structures (\texttt{a}) leads to a modest improvement in both pre-train/fine-tune and multi-headed training.
However, we find no additional benefit from training on the xTB labels of the test-set structures themselves, in contrast to the DFT case.
We attribute this to the weak alignment between xTB and CC labels, and hence to a reduction in the shared information content between the auxiliary and target tasks.

\begin{figure*}[!ht]
    \centering
    \includegraphics{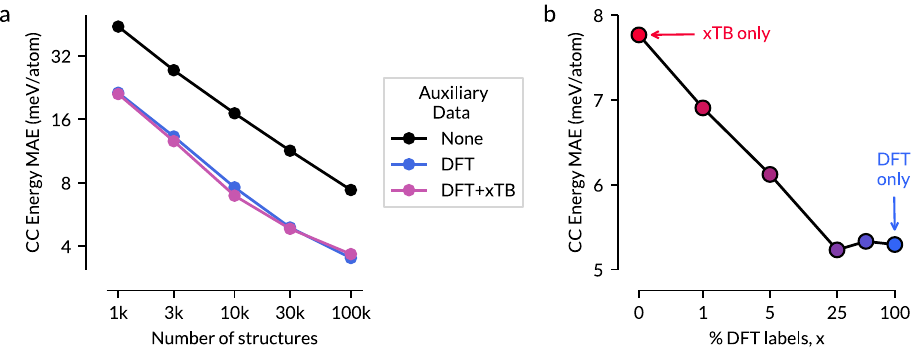}
    \caption{\label{fig:three-heads}%
        Training on three fidelities simultaneously.
        \textbf{(a)}
        Multi-headed training on combinations of three labelling methods, (CC; \texttt{a}), (DFT; \texttt{a}), and (xTB; \texttt{a}), using the same $n$ structures from each.
        The number of structures per source is varied along the $x$-axis; the error on the CC test set is plotted on the $y$-axis.
        Adding xTB as a third quantum-chemical method neither improves nor degrades the CC head's performance relative to training on CC and DFT alone.
        \textbf{(b)}
        Multi-headed training on 10k (CC; \texttt{a}) together with 100k auxiliary structures, of which $x\%$ are drawn from (DFT; \texttt{b}) and $(100{-}x)\%$ from (xTB; \texttt{c}).
    }
\end{figure*}

{\raggedright
    \subsection{Going Beyond Two Fidelities}
}

A practical advantage of the multi-headed architecture is its natural extensibility to more than two labelling methods, unlike the sequential pre-train/fine-tune approach.

We investigate this extensibility in Fig.~\ref{fig:three-heads}a by comparing multi-headed models trained using (i) DFT auxiliary labels (blue), and (ii) both xTB and DFT auxiliary labels (pink).
Encouragingly, we find that adding xTB as a third quantum-chemical method does not degrade the CC head's performance compared with training on CC and DFT alone.
This suggests that the compromise in backbone representations required to accommodate multiple labelling methods is a one-off cost, rather than scaling with the number of methods/heads.

DFT labels are significantly more expensive to generate than xTB labels.
In practice, therefore, one might wish to use a large amount of xTB-labelled auxiliary data, supplemented by a smaller amount of DFT-labelled auxiliary data.
To investigate the impact such a strategy has on the accuracy of the CC head, we train multi-headed models on 10k (CC; \texttt{a}) together with 100k auxiliary structures from a separate data split, of which the first $x\%$ are labelled with DFT, and the remaining $(100{-}x)\%$ are labelled with xTB.
We plot the CC-head's test-set error as a function of $x$ in Fig.~\ref{fig:three-heads}b.

Consistent with the pre-train/fine-tune experiments, a pure DFT auxiliary dataset ($x = 100$) yields larger gains on the CC task than a pure xTB dataset ($x = 0$).
However, the DFT fraction can be reduced to 25\% (25k DFT + 75k xTB) without significant loss in accuracy on the CC task, demonstrating that the performance gain from DFT labels is retained even when they are outnumbered by cheaper xTB labels.
Given the far lower computational cost of xTB relative to DFT, multi-headed training can therefore reduce the expense of auxiliary data generation without compromising target-task accuracy.

\section{Practical Recommendations}

Based on the preceding results, we offer the following recommendations for practitioners seeking to target a high-fidelity, energy-only labelling method (e.g., CC) via multi-fidelity training.

Pre-training on low-fidelity labels substantially improves downstream performance.
The pre-training stage should include both energy and force labels (Fig.~\ref{fig:ablation-forces}).
During fine-tuning, it is essential to use a small learning rate (Fig.~\ref{fig:repr-lr}) and to optimise all model parameters rather than just the readout head (Fig.~\ref{fig:repr:backbone}).

Multi-headed training provides a practical way to scale to more than two labelling methods simultaneously.
Practitioners can minimise computational cost by using predominantly cheaper auxiliary labels without significant loss in accuracy on the target fidelity (Fig.~\ref{fig:three-heads}).
The optimal range of relative sampling rates is broad; equal probability of sampling each dataset is near-optimal across all dataset sizes tested (Fig.~\ref{fig:mh-sampling-rate}).
When only two labelling methods are available, the pre-train/fine-tune approach typically achieves higher accuracy (Fig.~\ref{fig:mh-mace-vs-allegro}).

If a high-fidelity dataset is already available, the simplest starting point is to pre-train on the same structures labelled by a cheaper method before fine-tuning on the target labels.
Where possible, however, both strategies benefit from using distinct structures for the different fidelities (Fig.~\ref{fig:ablation-variety}).

The choice of low-fidelity method for pre-training depends on the total computational budget.
In the low-budget regime, the most effective strategy is to pre-train on abundant low-fidelity labels before fine-tuning on small amounts of high-fidelity labels.
At higher budgets, a more expensive method with better alignment to the target becomes preferable (Fig.~\ref{fig:cost-curve-xtb-dft}).

\section{Conclusion}

This study investigated multi-fidelity training strategies for MLFFs through systematic ablations, comparing pre-training/fine-tuning and multi-headed approaches to identify the mechanisms underlying positive transfer from low- to high-fidelity targets.

Pre-training on low-fidelity labels (DFT or xTB) before fine-tuning on CC energies consistently yields positive transfer, with gains most pronounced in the low-data regime (Fig.~\ref{fig:amount}).
The mechanism is indirect: models that are more accurate on the pre-training task produce better internal representations of local chemical environments, which in turn yield better fine-tuned performance.
This manifests as a log-log linear relationship between pre-training and fine-tuning accuracies that holds across amounts of fine-tuning data, model sizes, model architectures, and labelling methods (Figs.~\ref{fig:amount},~\ref{fig:dft-vs-xtb}c and~\ref{fig:allegro-vs-mace}b).
The transfer efficiency depends on the fidelity of the pre-training labels (Fig.~\ref{fig:dft-vs-xtb}a), the model's size (Fig.~\ref{fig:dft-vs-xtb}c), and, critically, on the inclusion of force labels (Fig.~\ref{fig:ablation-forces}).

Multi-headed models, which learn from multiple labelling methods simultaneously through a shared backbone, build method-independent representations that allow consistent positive transfer from the auxiliary heads to the target head (Fig.~\ref{fig:mh-mace-vs-allegro}).
Between the accuracies of the heads we also observed a log-log linear relationship (Fig.~\ref{fig:mh-val-aux-vs-cc}).
However, these shared representations entail a compromise: requiring the backbone to serve multiple heads simultaneously limits its specialisation, resulting in modestly lower accuracy than sequential pre-training/fine-tuning in many cases.
This trade-off is offset by practical advantages: multi-headed training naturally extends to more than two labelling methods.
Importantly, we see no further accuracy degradation due to the lack of backbone specialisation when moving from two to three heads.
We find that expensive DFT labels can be partially replaced by computationally cheaper xTB labels with minimal loss in accuracy on the CC task (Fig.~\ref{fig:three-heads}).
Positive transfer does not require structural overlap between datasets (Fig.~\ref{fig:ablation-variety}), making multi-headed training a robust and practical strategy when matched labels are unavailable.

This work compared two common multi-fidelity strategies using three quantum-chemical methods on a single dataset of small organic molecules.
Extending the analysis to broader chemical domains, such as metallic systems, ionic liquids, or biomolecules, would test how well these findings generalise.
A systematic comparison with $\Delta$-learning, which requires the base method at inference time but can be highly data-efficient, would further clarify the trade-offs among multi-fidelity approaches; connections to meta-learning also merit investigation.\cite{Allen2024Learning}
Multi-fidelity benchmark datasets spanning diverse chemical spaces and evaluating multiple approaches on a common footing would accelerate progress towards cost-efficient universal MLFFs.
Finally, this study assessed model quality through test-set accuracy alone; physical validation through molecular dynamics simulations and comparison to experimental observables remains essential for evaluating MLFFs in practice.\cite{Poltavsky2025CrashAnalysis,Poltavsky2025CrashMD,Simm2025SimPoly}

\section*{Code and Data Availability}

Code and data are available on \underline{\href{https://github.com/microsoft/multi-fidelity-training-mlff}{GitHub}}
and \underline{\href{https://huggingface.co/datasets/microsoft/multi-fidelity-training-mlff}{HuggingFace}}, respectively.

\section*{Acknowledgements}

We thank Michael Gastegger, Rianne van den Berg, and Chin-Wei Huang for their valuable feedback on the manuscript,
and Wenlin Chen, Anna Kuzina, and Yicheng Chen for their contributions to the codebase.
This research was fully funded by Microsoft Research.

\bibliography{main}

\onecolumn
\clearpage
\newgeometry{top=1in, bottom=1in, left=1.5in, right=1.5in}

\section{Appendix}

\setcounter{section}{0}
\setcounter{table}{0}
\setcounter{figure}{0}
\setcounter{page}{1}

\renewcommand{\thesection}{\Alph{section}}

\renewcommand{\thepage}{S\arabic{page}}
\renewcommand{\theequation}{S\arabic{equation}}
\renewcommand{\thefigure}{S\arabic{figure}}
\renewcommand{\thetable}{S\arabic{table}}

\subsection{ANI-1ccx Structures}

\begin{figure}[ht]
    \centering
    \includegraphics[width=0.8\linewidth]{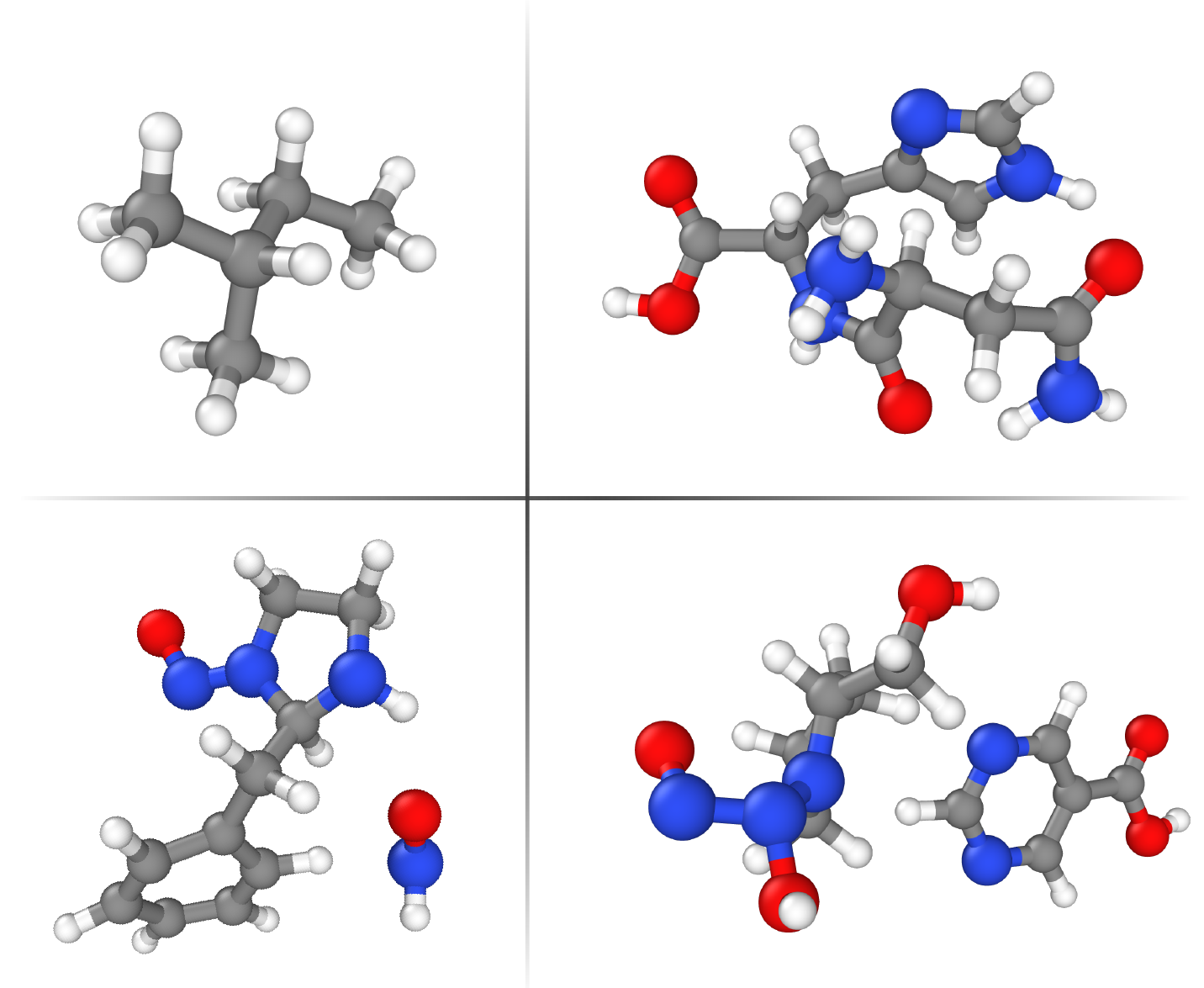}
    \caption{\label{fig:ani}
        Representative molecular structures from the ANI-1ccx dataset
        which encompasses both low- and high-energy conformations spanning various chemical compositions.\cite{Smith-20-05}
        Chemical structures are composed of carbon (grey), hydrogen (white), nitrogen (blue), and oxygen (red).
    }
\end{figure}

\subsection{Energy Alignment}

\begin{table}[H]
    \centering
    \begin{tabular}{rdddd}
        \toprule
            & \multicolumn{1}{c}{H} & \multicolumn{1}{c}{C} & \multicolumn{1}{c}{N} & \multicolumn{1}{c}{O} \\
        \midrule
        CC  & -16.57                & -1035.16              & -1487.48              & -2045.12              \\
        DFT & -16.51                & -1036.34              & -1488.50              & -2045.97              \\
        xTB & -13.98                & -57.96                & -78.66                & -110.22               \\
        \bottomrule
    \end{tabular}
    \caption{
        Per-element energy contributions (in eV) for the three quantum-chemical methods used in this work, calculated from the entire ANI-1ccx dataset using linear least-squares regression.
    }
    \label{tab:e0s}
\end{table}

\clearpage
\subsection{Comparison with ANI}

\begin{table*}[h]
    \centering
    \begin{tabular}{@{}cccc@{}}
        \toprule
        Architecture & \makecell{Direct                  \\ CC training} & \makecell{DFT pre-trained \\ CC fine-tuned} & \makecell{Positive \\ transfer} \\
        \midrule
        \makecell{ANI                                    \\ (Smith \textit{et al.}\cite{Smith-19-07})}
                     & \makecell{\qty{2.25}{\kcalpermol} \\ (500k CC)}
                     & \makecell{\qty{1.78}{\kcalpermol} \\ (5M DFT $\rightarrow$ 500k CC)}
                     & \qty{20}{\percent}                \\[4pt]
        \makecell{MACE                                   \\ (this work)}
                     & \makecell{\qty{2.23}{\kcalpermol} \\ (100k CC)}
                     & \makecell{\qty{0.62}{\kcalpermol} \\ (174k DFT $\rightarrow$ 100k CC)}
                     & \qty{72}{\percent}                \\
        \bottomrule
    \end{tabular}
    \caption{
        Comparison of direct CC training and DFT pre-trained/CC fine-tuned energy MAEs for the ANI and MACE architectures on the ANI-1ccx dataset.
        Positive transfer is the relative reduction in MAE as a result of pre-training.
        Exact comparison is not possible due to differences in train/test splits and amounts of data used; the numbers reported here are the best available for each architecture.
    }
    \label{tab:ani-vs-mace}
\end{table*}

\clearpage
\subsection{Model Capacity}
\label{sec:model_capacity}

\begin{figure}[ht]
    \centering
    \includegraphics{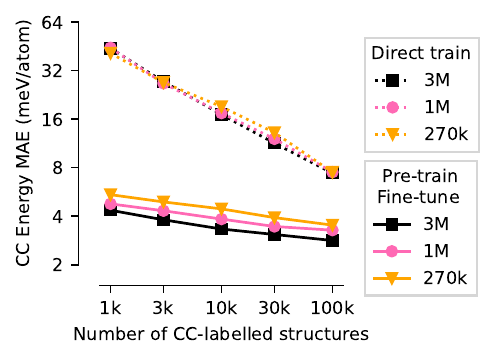}
    \caption{\label{fig:ablation-size}%
        Effect of model size on fine-tuned performance.
        Dashed lines: direct training on (CC; \texttt{a}).
        Solid lines: pre-training on 45k structures from (DFT; \texttt{b}) followed by fine-tuning on (CC; \texttt{a}).
    }
\end{figure}

To assess the role of model capacity in low-fidelity pre-training, we compare MACE models with 270k, 1M, and 3M parameters (corresponding to 64, 128, and 256 hidden channels, respectively).
When trained directly on (CC; \texttt{a}), all three model sizes perform similarly: each has sufficient capacity to fit 100k energy labels (dashed lines in Fig.~\ref{fig:ablation-size}; all models achieve training set MAEs of $\lesssim$\,\qty{1.5}{\mevpatom}).
After pre-training on 45k structures from (DFT; \texttt{b}) and fine-tuning on (CC; \texttt{a}), however, larger models show greater accuracy gains (solid lines in Fig.~\ref{fig:ablation-size}).
This difference traces back to the pre-training stage: the largest model achieves an MAE of \qty{3.1}{\mevpatom} on the DFT validation set, compared to \qty{7.2}{\mevpatom} for the smallest.
Because force labels carry substantially more information per structure than energy labels alone---each structure contributes $3N_s$ force components versus a single energy---smaller models lack the capacity to fully exploit this information, limiting the quality of their pre-trained representations.
Concretely, the force MAEs of the pre-trained models on the training set are 48, 31, and \qty{18}{\mevpang} for the small, medium, and large models, respectively.

\clearpage
\subsection{Learning Rate During Fine-tuning}
\label{sec:lr}

\begin{figure}[H]
    \centering
    \includegraphics{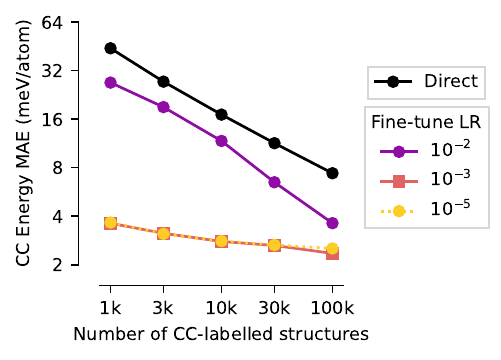}
    \caption{\label{fig:repr-lr}%
        Effect of the fine-tuning learning rate on model accuracy.
        Models are pre-trained on 90k structures from (DFT; \texttt{b}) and fine-tuned on increasing amounts of (CC; \texttt{a}) at different learning rates.
        The black line shows direct training on (CC; \texttt{a}) for reference.
    }
\end{figure}

We investigated the effect of the fine-tuning learning rate on predictive accuracy (Fig.~\ref{fig:repr-lr}).
Using the same learning rate as during pre-training ($10^{-2}$) yields only marginal improvement over direct training, likely due to catastrophic forgetting.
Reducing the learning rate to $10^{-3}$ recovers the full benefit of pre-training, while further reduction to $10^{-5}$ produces no additional gain.
This suggests that the loss landscape during fine-tuning is smooth enough to avoid entrapment in local minima at small learning rates.
We therefore use a learning rate of $10^{-3}$ for fine-tuning throughout this work.

\subsection{Cost-Error Analysis}

\begin{figure}[H]
    \centering
    \includegraphics{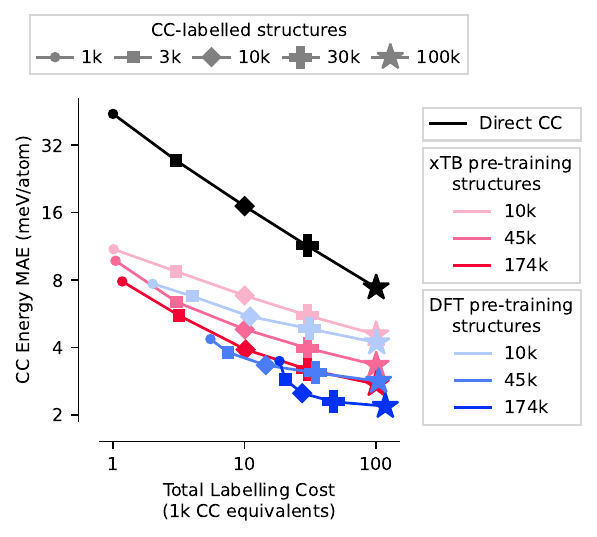}
    \caption{\label{fig:cost-curve-xtb-dft}%
        Comparing total cost of all labels used during both pre-training and fine-tuning (x-axis) against test-set CC energy MAE (y-axis) for various amounts of pre-training (shade) and fine-tuning (marker) data, and for both DFT (blue) and xTB (red) auxiliary labels.
        Both xTB and DFT pre-training strategies appear on the resulting cost-error Pareto front.
    }
\end{figure}

\clearpage
\subsection{xTB Auxiliary Experiments}

The following figures reproduce the DFT-based analyses from the main text using xTB as the auxiliary labelling method.

\begin{figure}[H]
    \centering
    \includegraphics{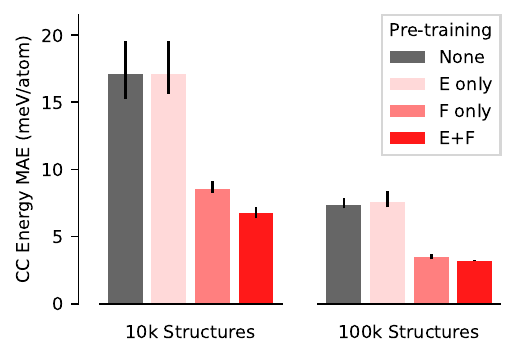}
    \caption{\label{fig:ablation-forces-xtb}%
        Effect of energy and force labels during xTB pre-training.
        Bars show the MAE on the CC test set; error bars span the min--max range.
        Left and right panels correspond to 10k and 100k training structures, respectively.
        Grey bars: direct training on (CC; \texttt{a}).
        Red bars: pre-training on (xTB; \texttt{a}) followed by fine-tuning on (CC; \texttt{a}).
        Shading indicates the label type used during pre-training: light for energies only, medium for forces only, dark for both.
    }
\end{figure}

\begin{figure}[H]
    \centering
    \includegraphics{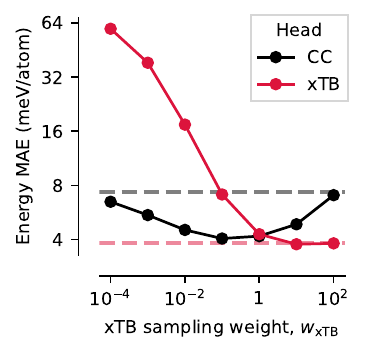}
    \caption{\label{fig:mh-sampling-rate-xtb}%
        Effect of the xTB sampling weight $w_\text{xTB}$ on the CC energy MAE on the test set after multi-headed training on (CC; \texttt{a}) and (xTB; \texttt{b}).
        Dashed lines indicate the MAE of single-headed models trained on CC (black) or xTB (blue) data alone.
    }
\end{figure}

\end{document}